\documentclass[reprint,prap,amsmath,amssymb,showpacs,twocolumn]{revtex4-1}
\usepackage{amssymb,amsmath,amsfonts,placeins,psfrag,graphicx,epsfig,mathtools}
\usepackage{color}
\usepackage[hidelinks]{hyperref}
\newcommand{\dd}{\partial}

\newcommand{\beq}{\begin{equation}}
\newcommand{\eeq}{\end{equation}}

\newcommand{\re}{\mbox{Re}}

\newcommand{\enquote}[1]{``#1''}

\begin{document}
\title{Large Q factor with very small Whispering Gallery Modes resonators} 
\author{Nirmalendu Acharyya$^{1,2}$}
\email{nirmalendu.acharyya@mbi-berlin.de}
\author{Gregory Kozyreff$^{2}$}
\email{gkozyref@ulb.ac.be}
\affiliation{$^1$Max-Born-Institut f\"ur Nichtlineare Optik und Kurzzeitspektroskopie, D-12489 Berlin, Germany.\\
$^2$Optique Nonlin\'eaire Th\'eorique, Universit\'e libre de Bruxelles (U.L.B.), CP 231, Campus de la Plaine, 1050 Bruxelles, Belgium.}
\date{\today}

\begin{abstract}
Efficient micro-resonators simultaneously require a large quality factor $Q$ and a small volume $V$. However, the former is ultimately limited by bending losses, the unavoidable radiation of energy of a wave upon changing direction of propagation. Such bending losses increase exponentially as $V$ decreases and eventually result in a drop of $Q$. Therefore, circular cavities are generally designed with radii that are much larger than the optical wavelength. The same leakage of energy by radiation limits the sharpness of bends in photonic integrated circuits. In this article, we present a way to reduce bending losses in circular micro-resonators. The proposed scheme consists of one or more external dielectric rings that are concentric with the cavity. These rings alter the field outside the cavity where radial oscillations set in, and thus control the far field radiation. As a result, the $Q$ factor can be increased by several orders of magnitude while keeping a small cavity volume.
\end{abstract}

\maketitle

\section{Introduction}
\begin{figure}
\flushleft \quad\includegraphics[width=7.5cm]{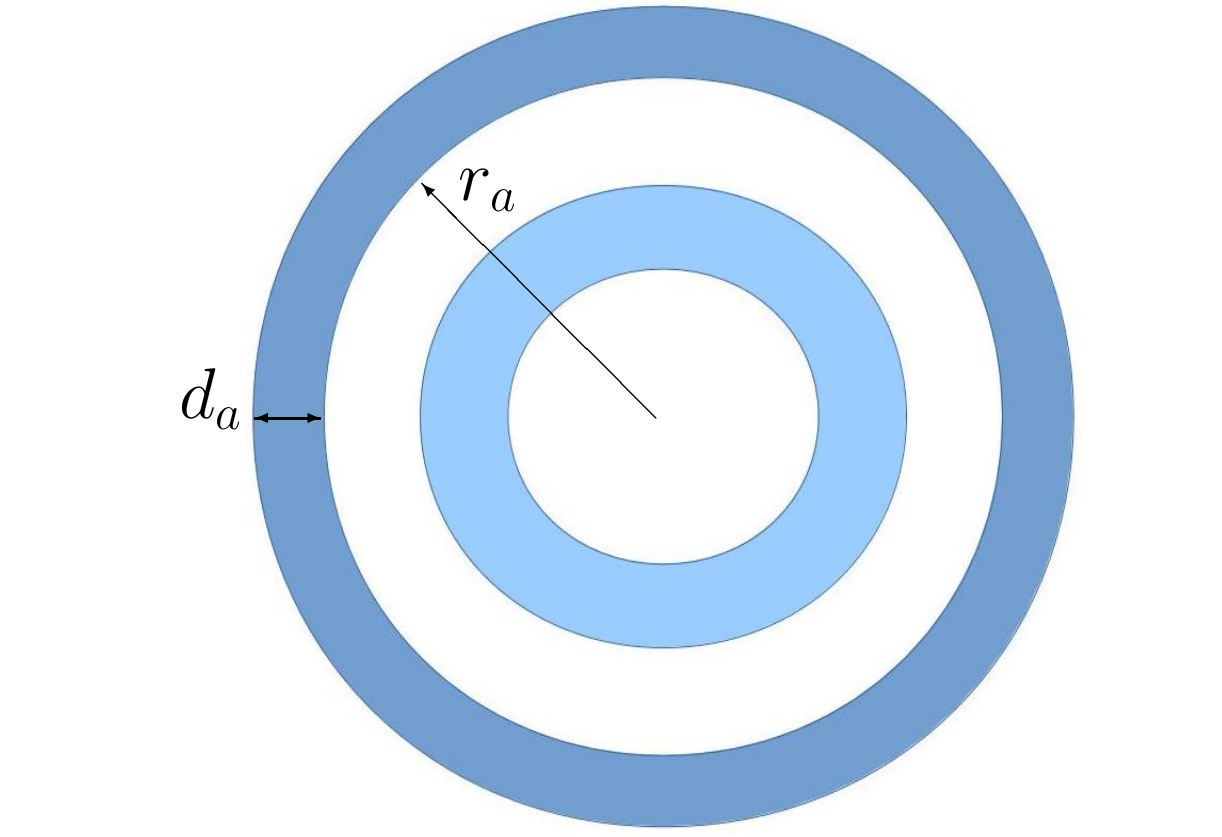}
\includegraphics[width=8cm]{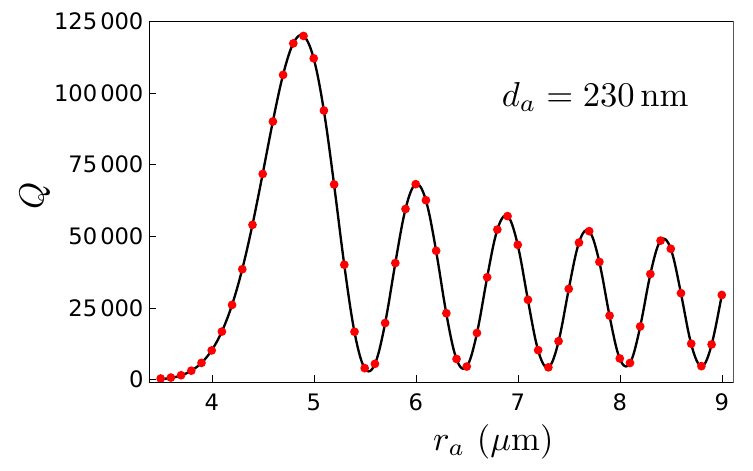}
\includegraphics[width=8cm]{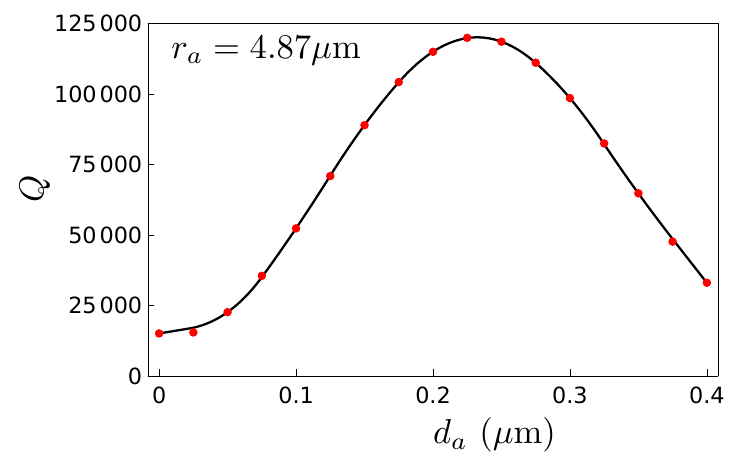}
\caption{Radiation quality factor in the presence of a single outer shell for a cavity of  radii $r_1=2.5\mu$m and $r_2=3.2\mu$m. Refractive index in the cavity and shell: $n=1.65$. Other regions: $n=1$. We consider a TE mode with $\nu= 22$ ($\lambda \sim 1.26 \mu m$.)  Without shell, $Q=Q^{c} \approx 15000$. Red dots:  numerical solution of the full characteristic Eq.~(\ref{eq:char1}). Solid lines: Eq.~(\ref{eq:Q}). }
\label{figoneshell}
\end{figure}

\begin{figure*}
\centering
{\includegraphics[]{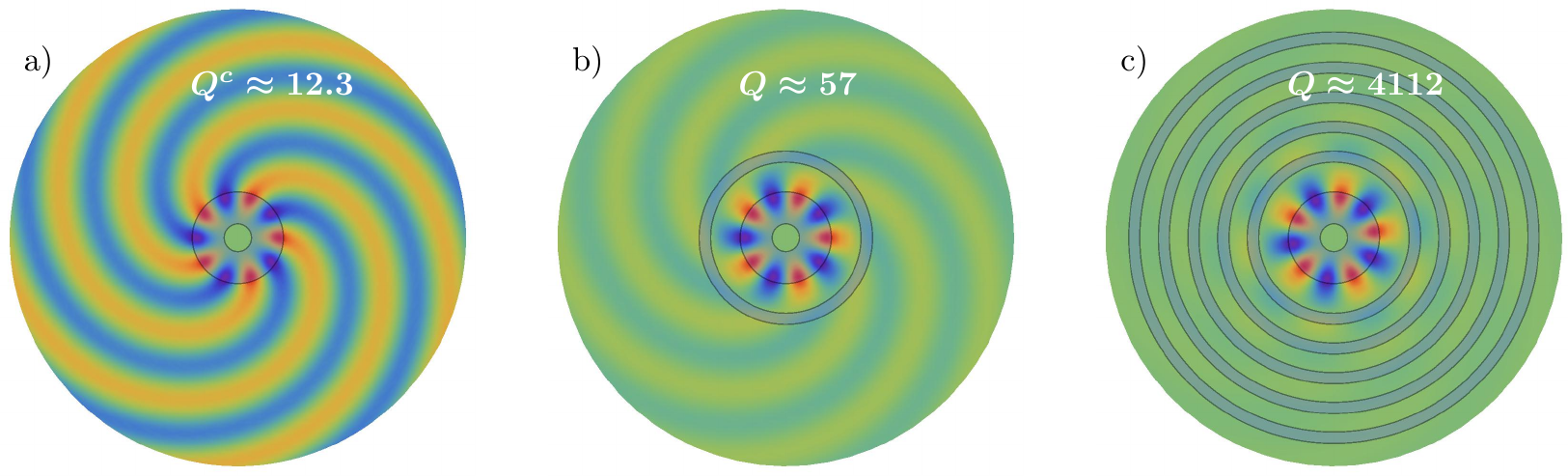}}
\caption{Field distribution $\re(E_z)$ for a TE mode with $\nu=5$ (see text). a: bare cavity; b: radiation with a single shell; c:~radiation with five shells. The cavity outer radius is $1\mu$m.}
\label{fignu5}
\end{figure*}

Whispering gallery mode (WGM) resonators occupy a place of choice in experiments and devices where enhanced light-matter interaction is required. The reason of their interest in both fundamental and applied research is that WGM can have a very long lifetime, measured by the quality factor $Q$, while occupying a small volume $V$~\cite{Matsko-2006,Ilchenko-review2}. This makes them extremely sensitive to changes in their environment~\cite{Vollmer-2012,Bogaerts-2012,Foreman-2015} and, hence, especially interesting as biosensors~\cite{Vollmer-2002,Arnold-2003,DeVos-2007,Washburn-2010,Iqbal-2010,Luchansky-2011,Washburn-2016}. Indeed, through linear processes, detection of single nanoparticles or molecules the size of a protein or a virus was experimentally demonstrated~\cite{Vollmer-2008,Lu-2011,He-2011}. Furthermore, similar detection limit in terms of mass has been reached, but with a set of molecules of much lower molecular weight ($<500$ Da) using nonlinear optics~\cite{Dominguez-2011,Kozyreff-2011}. Next to sensing, WGM are found to be particularly efficient nonlinear optical sources:  lasers~\cite{McCall-1992,Sandoghdar-1996,Srinivasan-2006}, optical parametric oscillators~\cite{Kippenberg-2004}, second-harmonic~\cite{Ilchenko-2004},  Raman~\cite{Lin-1994} and third-harmonic sources~\cite{Carmon-2007} and, in recent years, frequency combs~\cite{DelHaye-2007,Savchenkov-2008,Chembo-2010,Yi-2015}. Regarding phase-matching, it was realized~\cite{Kozyreff-2008,Jouravlev-2004}, and later experimentally confirmed~\cite{Furst-2010,Furst-2010b,Furst-2011}, that the usual conservation of linear momentum gives way to conservation of angular momentum in WGM resonators and that the laws of composition of angular momentum in quantum mechanics apply. Finally, WGM resonators can be used to investigate quantum cavity electrodynamics in the strong coupling regime~\cite{Vernooy-1998,Peter-2005,Aoki-2006} and   hold potential as high-quality optical quantum sources~\cite{Fortsch-2013}. In relation to linear light-matter interaction, an appropriate figure of merit is the Purcell factor $\mathcal P=6\pi Q/(k_r^3V)$, where $k_r$ is the real part of the wavenumber in vacuum. Indeed, $\mathcal P$ characterises cavity-induced changes not only in emission but also in scattering by particles, both in the quantum and classical limits~\cite{Ozdemir-2011}. As a general rule, therefore, the  optimisation of WGM resonators rests on maximising $Q$ while keeping $V$ as small as possible.

In large cavities, the quality factor is limited by parasitic absorption and scattering. By minimizing these losses with state-of-the fabrication techniques, $Q$ in the order of $10^{11}$ have been demonstrated~\cite{Savchenkov-2007b} while values in excess of  $10^6$ have been reached in various photonic integrated platforms~\cite{Tien-2011,Spencer-2014,Hausmann-2014,Su-2018,Shitikov-2018,Yang-2018}. However, this is no longer true as the cavity radius $R$ becomes smaller than about ten wavelengths. Indeed, radiation losses associated to the bending of light trajectories increases exponentially with curvature. They are therefore the main physical obstacle to reduce $V$ while preserving $Q$. 

In this article, we propose a general procedure to reduce bending losses, which can readily be implemented with existing technologies. It consists of surrounding the circular cavity with properly designed concentric dielectric shells. These give control to the far-field amplitude and, hence, the radiation losses of WGM. The more external shells there is in the external structure, the stronger the reduction of bending losses can be obtained, with no apparent limit. As an example, we show an increase of the $Q$ factor from 15\,000 to 125\,000, 600\,000 and 2.5 10$^6$ with one, two or three external rings, respectively (here, and from now on, we restrict our attention to the radiative part of the $Q$ factor.) In another, extreme case, we consider a WGM with orbital number $\nu=5$ and vacuum wavelength $\lambda=1.45\mu$m in a cavity with $R=1\mu$m only. Starting from an initial quality factor $Q=12.3$, we obtain $Q\approx4699$ by adding five external shells, while keeping the WGM entirely within the cavity.  The strongest suppression of radiation is generally obtained when the innermost external shell is at the transition where the WGM field switches between exponential (near field) and undulatory (far-field) behaviour. Conversely, if desired, the external structure can drastically enhance radiation losses and decrease $Q$. Hints that radiation losses could be controlled by such a structure were found by an analytical theory in the limit of large $R$ with a single outer shell~\cite{Kozyreff-2016}. Here, we provide a complete theory, that remains valid for small $R$ and which applies to any number of external shells. 
Previous authors have considered geometries that appear at first sight similar to what is considered here~\cite{Li-2009,Cai-2015,Malmir-2016,Chamorro-Posada-2017,Chamorro-Posada-2018} (Note for proper comparison that the field circulates in the external ring in Ref.~\cite{Malmir-2016}.) However, the present scheme is different in some fundamental aspects, which explains that the enhancement in $Q$ reported here is orders of magnitudes larger that in previous works: An enhancement factor at most between 2 and 10 in \cite{Cai-2015,Chamorro-Posada-2017} compared to  more than 350 in some examples discussed here. In Refs.~\cite{Li-2009,Cai-2015,Malmir-2016,Chamorro-Posada-2017,Chamorro-Posada-2018} the cavity is coupled to an external, curved waveguide, which effectively increases the radius of the ring. By contrast, the external rings considered in this Letter are too thin to act as waveguide and the field remains entirely confined inside the cavity. When it contains many shells, the present configuration becomes closer in spirit to Bragg fibres~\cite{Yeh-1978,Xu-2000}. However, we consider photons that circulate in a cavity rather than propagate along the axis of a Bragg fibre. In addition, in the optimal configuration, the first ring is always in the evanescent zone of the cavity and not in the radiation zone, a consideration that is evidently absent in the study of Bragg fibres.

\section{Theory}
An analytical understanding of the radiation of a WGM resonator embedded in such a dielectric `sarcophagus' can be obtained in 2D. In this framework, the knowledge of the electromagnetic field is entirely encoded in just one of its component, $\psi=E_z$ for transverse electric (TE) modes or $\psi=H_z$ for transverse magnetic (TM) modes. In an annulus defined by $r_{j-1}<r<r_j$, of refractive index $n_j$, the general form of $\psi$ is
\beq
\psi = \left[a_j J_\nu(n_j k r)+ b_j Y_\nu(n_j k r)\right]e^{i\nu\theta-ikc t}
\label{eq:JY}
\eeq
where $r$ and $\theta$ are the usual polar coordinates, $c$ is the speed of light in vacuum, $J_\nu, Y_\nu$ are Bessel functions of the first and second kind~\cite{Abramowitz} and $k$ is the complex wave number. At surfaces of discontinuity of the refractive index, both $\psi$ and either $\dd\psi/\dd r$ (TE) or $n^{-2}\dd\psi/\dd r$ (TM) are continuous. These continuity relations can expressed as 
\beq
\begin{pmatrix}a_{j-1}\\b_{j-1}\end{pmatrix} = S_j\begin{pmatrix}a_{j}\\b_{j}\end{pmatrix}.
\eeq
where the $S_j$ are $2\times2$ matrices containing combinations of Bessel functions evaluated at appropriate interfaces (see Supplementary Information.) By iterating the process, one may link the innermost and outermost coefficients of Eq.~(\ref{eq:JY}), giving
\beq
\begin{pmatrix}a_0\\0\end{pmatrix} = S(k)\begin{pmatrix}a_{N}\\ i a_{N}\end{pmatrix},
\label{eq:cont1}
\eeq
with $S=S_1S_2\ldots S_N$. Above, we have imposed constraints on the combinations of Bessel functions near the origin and in the outermost region, namely to avoid  divergence as $r\to0$ and impose proper radiation condition in the far field. The second component of Eq.~(\ref{eq:cont1}) directly yields the characteristic equation:
\beq
S_{21}(k)+i S_{22} (k)=0.
\label{eq:char1}
\eeq
It has complex roots of the form $k=k_r-i k_i$, from which the quality factor can be deduced as $Q=k_r/(2k_i)$. To study $Q$ by direct resolution of Eq.~(\ref{eq:char1}) for each choice of geometrical parameter set $\{r_j\}$ rapidly becomes intractable as the number of outer rings increases. However, our aim here is to study the effect of the external structure on the radiation properties of the internal one, \emph{i.e.} to compare $k$ with the complex wave number  $k^{c}=k^{c}_r-ik^{c}_i$ of the bare cavity. The latter is easier to compute, as it solves a simpler equation. Moreover, in the situations of interest,  $k^{c}_i$ is by hypothesis not so small that it requires special numerical care. As the innermost layers of the whole guiding structures make up the bare cavity, we may write $S$ in Eq.~(\ref{eq:cont1}) as $S=S^{c}S^{s}$, where $S^{c}$ and $S^{s}$ correspond to the cavity and the radiation shielding structure, respectively. By analogy with the above, $k^{c}$ satisfies the simpler equation
\beq
S^{c}_{21}(k^{c})+i S^{c}_{22} (k^{c})=0,
\label{eq:char2}
\eeq
With $k^{c}_i\ll k^{c}_r$, it must either be that, to leading order, $|S^{c}_{21}(k^{c}_r)|\ll1$ and $S^{c}_{22}(k^{c}_r)=0$ or that $S^{c}_{21}(k^{c}_r)=0$ and $|S^{c}_{22}(k^{c}_r)|\ll1$. If the cavity is a simple disk, then it is easy to check that $S^{c}_{22}(k^{c}_r)=0$ is precisely the sought-after characteristic equation. For more complicated geometries, we find that it is still so. This can be traced back to the fact that $S_{22}/S_{21}$ is on the order of $Y_\nu(n_1kr_1)/J_\nu(n_1kr_1)$ which is a rapidly increasing function of $\nu$~\cite{Abramowitz}. Hence, a complex resonance of the bare cavity is approximately given by
\begin{align}
S^{c}_{22}(k^{c}_r)&=0, &k^{c}_i \approx - S^{c}_{21} (k^{c}_r)/S'^{c}_{22} (k^{c}_r),
\label{sol:cav}
\end{align}
where prime denotes derivative. In the situations of interest here, $\nu$ not being very large,  finding the root of $S^{c}_{22}(k_r)$ does not pose a numerical challenge
(for a large-$\nu$ treatment, see~\cite{Kozyreff-2016}.) Next, with $S=S^cS^s$, Eq.~(\ref{eq:char1}) yields
\beq
S^{c}_{21}\left(S^{s}_{11} +i S^{s}_{12} \right)+S^{c}_{22}\left(S^{s}_{21} +i S^{s}_{22} \right)=0.
\label{eq:char3}
\eeq
Focusing on the solutions that correspond to perturbed modes of the bare cavity, we note that $k=k_r^c+\Delta k$, with $|\Delta k|\ll k_r^c$. Expanding Eq.~(\ref{eq:char3}) near $k_r^c$ and exploiting Eq.~(\ref{sol:cav}), we obtain
\beq
\Delta k\approx -\left[\frac{S^{c}_{21}\left(S^{s}_{11} +i S^{s}_{12} \right)}{S'^{c}_{22}\left(S^{s}_{21} +i S^{s}_{22} \right)}\right] 
\approx -i k_i^c \left[\frac{ S^{s}_{11} +i S^{s}_{12} }{ S^{s}_{22}-iS^{s}_{21}  }\right],
\eeq
where the elements on the right hand side are evaluated at the known value $k_r^c$. Since $\Delta k= k_r-k_r^c-i k_i$, the change in quality factor that results from the external structure is found to be:
\beq
\left(\frac{Q}{Q^c}\right)^{-1}=
 \frac{k_i}{k_i^c} \approx \re\left[\frac{ S^{s}_{11} \left(k_r^c\right)+i S^{s}_{12} \left(k_r^c\right)}{ S^{s}_{22}\left(k_r^c\right)-iS^{s}_{21}\left(k_r^c\right)  }\right].
 \label{eq:Q}
 \eeq
The advantage of the above expression is that it explicitly yields the ratio $Q/Q^c$ without having to solve the characteristic equation of the complete geometry. It proves  to be extremely accurate for values of the orbital number $\nu>10$ (see Fig~\ref{figoneshell}.) Even for $\nu=5$ does it predict the enhancement of $Q$ to within fifteen percent. There has been a few previous works (see \cite{Kozyreff-2016} and references therein) to estimate the losses in the asymptotic cases of large circular orbital number $\nu$. However, to our knowledge,  an analytic formula such as Eq.~(\ref{eq:Q}) which is nearly exact has never been presented. The provided formula is very useful, particularly to optimise a shield with many shells where a numerical resolution of the characteristic equation is laborious. It enables us to circumvent the problems of solving the transcendental equation and merely requires evaluating the formula for various shield parameters. 

The above analysis suggests a simple, layer-by-layer, design strategy. Starting form the bare cavity, one first considers a single outside shell with inner and outer radii $r_a$ and $r_a'=r_a+d_a$. To optimize the right-hand-side of Eq.~(\ref{eq:Q}) with respect to only two parameters $r_a$ and $d_a$ is a straightforward matter. The greatest single-step enhancement is usually seen with this first outer shell. While several local maxima in the gain $Q/Q^{c}$ are found, see Fig.~\ref{figoneshell}, the global maximum is typically found near the turning point $n R$ where the spatial field distribution switches from exponential to oscillating. Subsequent improvements are then achieved by optimizing the parameters of a second shell structure, followed by a third one, \emph{etc}. To demonstrate this procedure, we first consider an Al$_2$O$_3$ ring cavity  (refractive index $n=1.65$) in air environment.  Al$_2$O$_3$ has been demonstrated to be an advantageous, CMOS compatible, host for rare-earth dopant and holds great potential to integrate micro-lasers in photonic platforms~\cite{Bernhardi-2010,Bernhardi-2011,Bernhardi-2012,Frankis-2018}. The bare ring cavity has  inner and outer radii given by $2.5$ and $3.2\mu$m, respectively. We focus on the fundamental radial TE mode with orbital number $\nu=22$, which corresponds to a wavelength $\lambda\approx1.26\mu$m, in the emission band of Yb. Without a dielectric sarcophagus, $Q^{c}\approx 15\,000$. With a single shell with parameters $(r_a,d_a)=(4.87,0.23)\mu$m, an eight-fold increase of $Q$ is obtained. Approximate four-fold increases are additionnally gained with a second and third shell with inner radii and thicknesses  $(r_b,d_b)=(5.63,0.21)\mu$m, and $(r_c,d_c)=(6.3,0.2)\mu$m, respectively, eventually raising $Q$ to $2.5\times 10^6$ (see Supplemental Information.) This last value is close to the current intrinsic limit of Al$_2$O$_3$. 
Interestingly, Fig.~\ref{figoneshell} indicates that $Q$ can also be significantly \emph{decreased} for other configurations; in that case, WGM radiation is enhanced by the external structure. 

As a second example, we consider a ring cavity with only 1$\mu$m outer radius and radial thickness $0.7\mu$m, operating at $\lambda\approx1.45\mu$m, that is $\nu=5$. Here, $Q^c=12.3$, which is unacceptably low compared to the state of the art. Adding five layers, of internal radii $(r_a,r_b,r_c,r_d,r_e)=(1.55, 2.22, 2.85,3.47, 4.07)\mu$m of respective thicknesses $(d_a,d_b,d_c,d_d,d_e)=(0.24, 0.23, 0.23,0.22, 0.22)\mu$m
leads to $Q=4699$, representing and enhancement by more than a factor 350. Fig.~\ref{fignu5} shows the change in radiation intensity and demonstrates that the mode energy stays confined in the central part of the structure. Further improvement can be obtained with additional layers. We note in this example that the optimal value of $d$ is close to $\lambda/4n$. It is expected that $d$ tends to that limit in the far-field, as the WGM locally tend to plane waves and the shield becomes equivalent to a Bragg reflector.

The above remark strongly suggests that the physical mechanism behind WGM radiation shielding is a kind of Bragg reflection by the external shells. Indeed, in the Supplementary Information, the shells are seen to nearly exactly contain a quarter of radial oscillation of $\psi$ in the optimal configuration, as with plane waves~\cite{Wolf-1980}. There are two differences with respect to standard Bragg reflection, however. Firstly, the radial oscillation are not sinusoidal but governed by Bessel functions. Consequently, the radiation shield is not periodic and its design rest on a formula like Eq.~(\ref{eq:Q}). Secondly, and more fundamentally, the efficiency of the shield critically depends on where it is located -a feature that is obviously absent from standard Bragg reflection. 
Whereas the problem of reflection of plane waves is invariant by translation, the WGM cavity introduces an absolute reference point on the radial axis. 

To illustrate this last point, let us resume the first example above, a bare  Al$_2$O$_3$ cavity of external radius $3.2\mu$m operating on the $\nu=22$ azimuthal mode, with $Q^c \approx 15000$. If one now encircles the cavity with two  external shells with  $(r_a,d_a)=(4.0,0.15)\mu$m and $(r_b,d_b)=(4.6,0.13)\mu$m, one obtains  $Q\approx 215$, approximately corresponding to a 70-fold \emph{enhancement} of the power radiated by the mode. Such a phenomenon can not be interpreted by simply picturing the external shells as a reflector. How the radial dependence of the field is affected by the shells is shown in the Supplemental Material.

\section{3D simulations}
\begin{figure}[h!]
\centering
{\includegraphics[width=7cm]{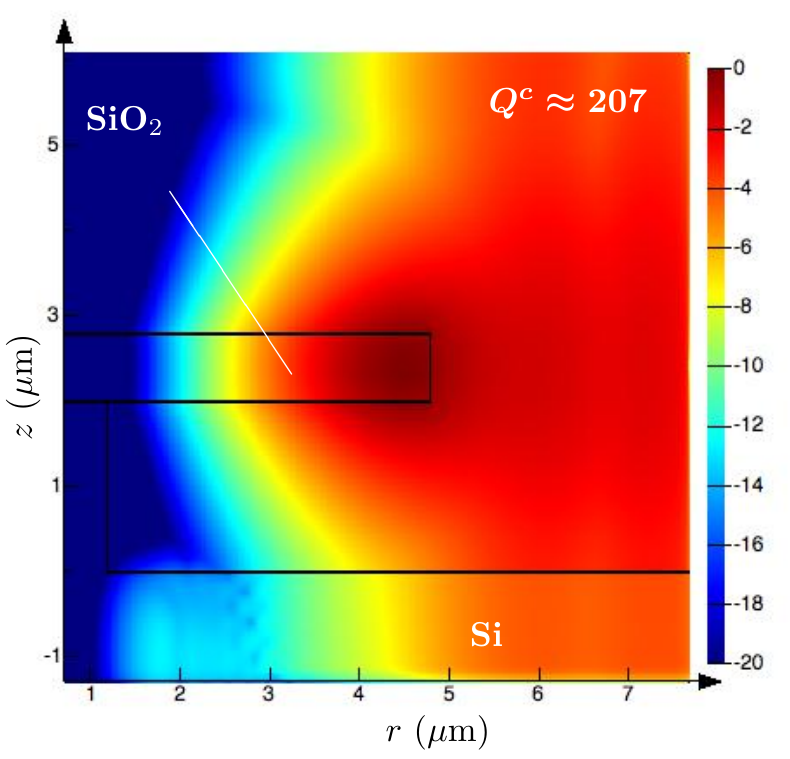}}
{\includegraphics[width=7cm]{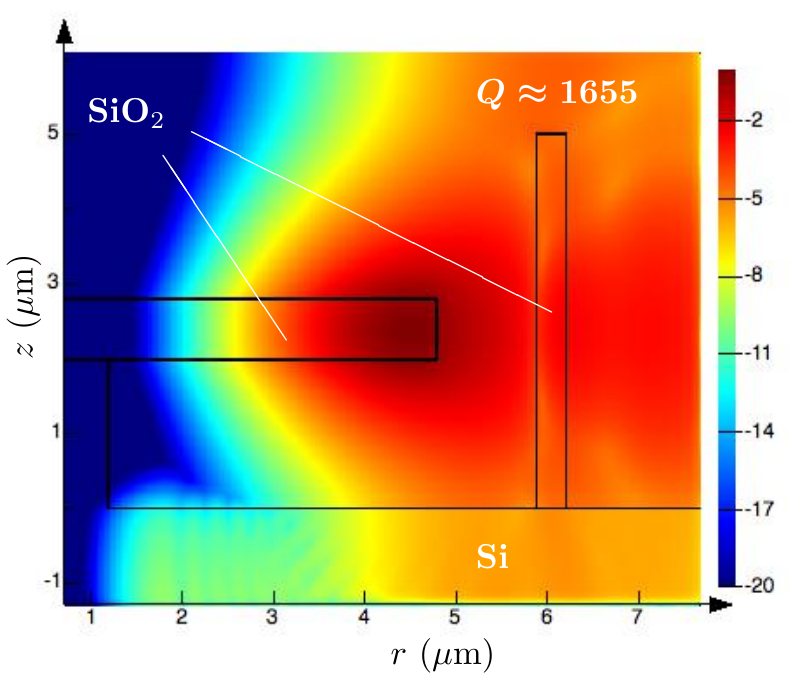}}
{\includegraphics[width=7cm]{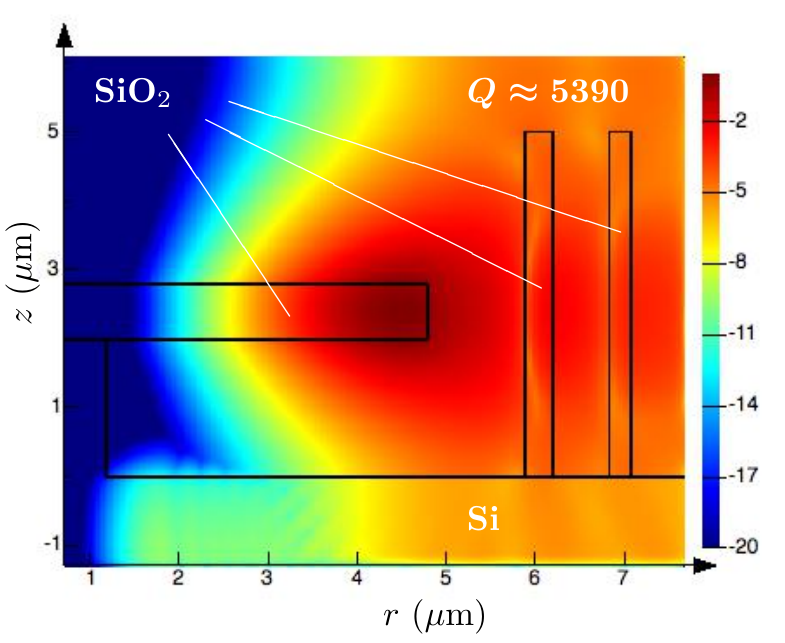}}
\caption{Mode distribution (logarithmic color scale) of a SiO$
_2$ disk cavity over a Si base.  Cavity radius and thickness: $4.78\mu$m and $0.8\mu$m, respectively.  First shield inner radius and thickness: $r_a=5.87\mu$m, $d_a=0.33\mu$m. Second shield: $r_b=6.84\mu$m, $d_b=0.23\mu$m. Height of both shields: $5\mu$m. Vacuum wavelength: $1.27\mu$m.
}
\label{fig:disk}
\end{figure}

It is intuitively clear that, as far as radial confinement is concerned, the 2D picture provides a faithful representation of WGM, even in 3D. Indeed the WGM on a sphere with orbital number $\nu$ have a radial dependence, and a characteristic equation, again controlled by Bessel functions, albeit of order $\nu+1/2$ instead of $\nu$~\cite{Matsko-2006}. The WGM on a sphere can thus be mapped onto those of a  an infinite cylinder and the two spectra coincide up to the transformation $\nu\to\nu+1/2$. On the other hand, spherical WGM can be strongly confined in the polar direction, with their intensity distribution confined in the immediate vicinity of the equator. Those particular WGM are almost unaffected if the sphere is truncated along parallel planes to the equator, which, in turn, is geometrically similar to a disk. Hence, we expect that  2D cylindrical WGM can serve as a reasonable qualitative model of  WGM in ring and disk cavities of finite vertical height and that our finding can be transposed there. To check this assertion, we have performed 3D simulations of SiO$_2$ disk cavities on a pilar over a Si foundation, as in Figure~\ref{fig:disk}. Such a configuration, or similar ones, can be made using  xenon difluoride (XeF$_2$) etching through a silicon substrate~\cite{Yang-2018}. Here again, we manage to obtain a nearly eight-fold enhancement with a single external shell. With a second external shell the initial $Q$ can be improved further, up to a factor 26. This confirms both the fact that the quality factor can be increased by the design of an external shell and the fact that further enhancement can be obtained by additional shell. Hence, the gain reported in these 3D simulation should by no means be considered as ultimate ones.

\begin{figure}
\centering
{\includegraphics[width=7cm]{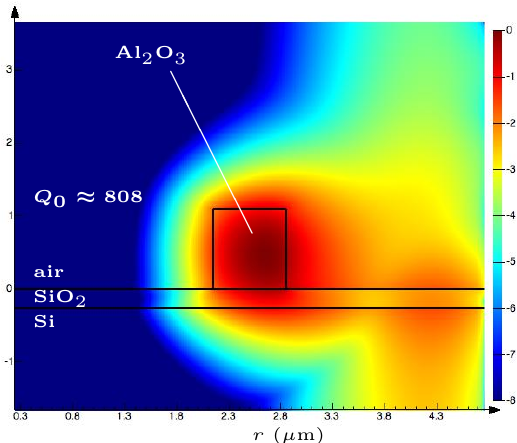}}
{\includegraphics[width=7cm,height=6.125cm]{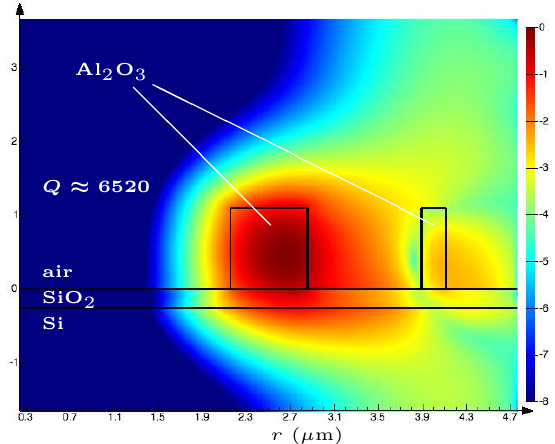}}
\caption{Cross-sectional field (logarithmic color scale) of a Al$_2$O$_3$
 ring made of ridge waveguides.  Cavity radius, thickness and height: $2.5\mu$m, $0.7\mu$m and $1.1 \mu$m, respectively. The base layer of SiO$_2$ is 0.25 $\mu$ thick. The shield has inner radius and thickness: $r_a=3.85\mu$m, $d_a=0.23\mu$m. Height of the shield: $1.1\mu$m. Vacuum wavelength: $1.27\mu$m.
}
\label{fig:ridgeWG}
\end{figure}

Finally, in Fig.~\ref{fig:ridgeWG}, we confirm the effect with a ridge waveguide ring cavity. Again, an eight-fold improvement of $Q$ is obtained with a  judiciously positioned external shell. Note that the vertical confinement renders possible the excitation of the cavity in the presence of the external shells. For instance, ridge-waveguide cavities can efficiently be excited with buried waveguides~\cite{Su-2018,Frankis-2018}.

\section{Conclusions}
In conclusion, we have devised a way to control one of the most basic limiting factor to WGM resonator performance. With existing fabrication techniques, radiation losses can in principle be reduced to any desired degree on integrated photonic platforms. In the ratio $Q/V$, the quality factor could thus be limited by material factors only and not by bending losses. This could pave the way to orders of magnitude improvement of  performances in laser operation, sensing or cavity quantum electrodynamics experiments based on WGM.

\acknowledgments
G.K. is a Research Associate of the Fonds de la Recherche Scientifique - FNRS (Belgium.) 
This research has received funding from the European Union's Seventh Programme for research, technological development and demonstration under grant agreement No 634928 (GLAM project, \url{http://www.glam-project.eu/}).

\newpage


%

\end{document}